\documentclass[12pt]{article}
\usepackage[utf8]{inputenc}
\usepackage[T1]{fontenc}
\usepackage{amsmath,amssymb} 
\usepackage{newtxtext,newtxmath}
\usepackage{graphicx}
\usepackage{enumitem}
\usepackage{longtable}
\usepackage{booktabs}
\usepackage{makecell}
\usepackage{listings}
\usepackage{enumitem}
\usepackage{array}
\usepackage{calc}
\usepackage{bm}
\usepackage{doi}
\usepackage{hyperref}
\pdfoutput=1
\title{Estimation of Industrial Heterogeneity from Maximum Entropy and Zonotopes Using the Enterprise Surveys}
\author{Tingyen Wang}
\date{}
\begin{document}
\maketitle
This study introduces a novel framework for estimating industrial
heterogeneity by integrating maximum entropy (ME) estimation of
production functions with zonotope-based measures. Traditional
production function estimations often rely on restrictive parametric
models, failing to capture firm behavior under uncertainty. This
research addresses these limitations by applying Hang K. Ryu's ME method
to estimate production functions using World Bank Enterprise Survey
(WBES) data from Bangladesh, Colombia, Egypt, and India. The study
normalizes entropy values to quantify heterogeneity and compares these
measures with a zonotope-based Gini index. Results demonstrate the ME
method's superiority in capturing nuanced, functional
heterogeneity often missed by traditional techniques. Furthermore, the
study incorporates a "Tangent Against Input Axes" method to dynamically
assess technical change within industries. By integrating information
theory with production economics, this unified framework quantifies
structural and functional differences across industries using firm-level
data, advancing both methodological and empirical understanding of
heterogeneity. A numerical simulation confirms the ME regression
functions can approximate actual industrial heterogeneity. The research
also highlights the superior ability of the ME method to provide a
precise and economically meaningful measure of industry heterogeneity,
particularly for longitudinal analyses.

\subsection*{1. Introduction}

\hspace*{2em}Traditional production function estimation has relied on parametric
models like Cobb-Douglas and translog, which, despite offering
foundational insights, impose restrictive functional forms and struggle
to reflect firm behavior under uncertainty. Especially when technical
change and firm heterogeneity are prominent, methods like OLS and
maximum likelihood often suffer from bias and misspecification due to
their inability to capture non-linearities and structural breaks.

\setlength{\parskip}{\baselineskip}\hspace*{2em}In response, entropy-based methods from information theory have emerged
as robust alternatives, successfully applied in fields like machine
learning and telecommunications. Though historically limited in
economics, recent studies show their promise in efficiency analysis and
firm-level performance. Grey system-based normalization techniques also
enhance robustness under uncertain or incomplete data. Together, these
approaches provide a flexible, information-theoretic framework for
evaluating industrial heterogeneity. Industrial
heterogeneity---differences in firm behavior, efficiency, and technology
within sectors---is key to understand productivity and competition.
Traditional methods yield useful insights but fall short in modeling the
diversity and uncertainty of modern industries. Measures like the Gini
coefficient from Zonotope quantify disparities but overlook functional
heterogeneity, prompting the need for more nuanced tools.

\hspace*{2em}Seminal works by Melitz (2003) and Syverson (2011) emphasize the role of
heterogeneity in shaping aggregate industry outcomes. Building on this,
our study applies Hang K. Ryu's maximum entropy method to estimate
production functions and normalize entropy values to measure
heterogeneity. Prior applications, such as Howitt and Msangi (2014) in
agriculture and Kayzel and Pijpers (2023) in network reconstruction,
demonstrate the method's versatility. Apeti and Ly (2024) showed how
power constraints differentially impact productivity across countries.
Entropy-based indices like ANE (El-Gamal \& Inanoglu, 2005) and Theil
entropy (Trushin \& Ugur, 2021) link dispersion to inefficiency,
survival, and innovation. Alternative approaches, including quantile
regression (e.g., Montresor \& Vezzani, 2015; Avenyo et al., 2021), also
underscore the importance of capturing heterogeneity in production
analysis. Empirical evidence from World Bank Enterprise Surveys (WBES)
further highlights firm-level variability.

\hspace*{2em}This study uses WBES data from five countries (Bangladesh, Colombia,
Egypt, India) and selected ISIC industries to compare entropy-based
measures with a zonotope-based Gini index following Dosi et al. (2016).
Results show the superiority of the maximum entropy method in capturing
nuanced, functional heterogeneity often missed by traditional
techniques. Overall, this research (1) introduces maximum entropy
estimation for production functions and (2) demonstrates how normalized
entropy values can serve as reliable metrics for industrial
heterogeneity. By integrating information theory with production
economics, we present a unified framework to quantify structural and
functional differences across industries using firm-level
data---advancing both methodological and empirical understanding of
heterogeneity.

\hspace*{2em}While this study introduces maximum entropy estimation for production
functions and demonstrates how normalized entropy values can serve as
metrics for industrial heterogeneity, it is important to acknowledge
that the illustrative examples presented rely on simplified assumptions,
particularly in the calculation of firm-level productivity, which are
discussed in more detail later in the paper.

\hspace*{2em}The remainder of this article is organized as follows. Section 2
outlines the introduction of WBES data and reviews relevant data
preprocessing. Section 3 presents the methodological design and details
the empirical procedures, including the normalization of entropy and the
use of the Gini coefficient from Zonotope, etc. Section 4 discusses the
empirical findings from World Bank data, comparing the maximum entropy
approach with Zonotope methods. Section 5 shows the experiment of
simulation to prove our findings. Finally, Section 6 concludes the study
and identifies potential directions for future research.
\subsection*{2. Data}
\hspace*{2em}This study utilizes firm-level data from the World Bank Enterprise
Surveys (WBES), a comprehensive dataset that provides detailed insights
into the business environment and firm performance across a wide range
of developing and emerging economies. The WBES collects standardized
information through stratified random sampling, ensuring
representativeness across regions, industries, and firm sizes within
each economy. The surveys cover private firms in manufacturing and
services sectors, excluding agriculture and extractive industries, and
include key variables such as sales, labor, capital, and intermediate
inputs, which are critical for estimating total factor productivity
(TFP).

\hspace*{2em}Following the methodology outlined in Francis et al. (2020), the data
preprocessing involves several steps to ensure consistency and
reliability for TFP estimation. First, all monetary values, originally
reported in nominal local currencies, are converted to U.S. dollars
using exchange rates from the World Bank's World Development Indicators.
These values are then deflated to real terms using the producer price
index (PPI) sourced from the International Monetary Fund (IMF), which
captures price changes from the perspective of domestic producers and is
deemed more suitable for firm-level output deflation than the broader
GDP deflator. Second, outliers are addressed by trimming extreme values
of key variables (e.g., sales, labor, and capital) to mitigate the
impact of measurement errors or reporting inconsistencies. Third, to
ensure comparability across economies and over time, the dataset is
pooled at the two-digit industry level based on the International
Standard Industrial Classification (ISIC), enabling robust estimation of
production functions. Missing data for critical variables are handled
through imputation techniques where feasible, following standard
practices to preserve sample size without introducing significant bias.

\hspace*{2em}The dataset is initially loaded using the pandas library, followed by a
preprocessing step that generates a subset of the data, a filtered
long-format dataset, and dictionaries mapping specific variables (e.g.,
(d2), (n7a), (n2a)). To align the subset with the filtered long-format
data, a merge is performed based on the variables ``country\_official'',
``year'', and ``isic'', assigning corresponding row indices from the
filtered dataset. Rows with missing indices are excluded to ensure data
consistency. Subsequently, the dataset is grouped by
``country\_official'' and ``isic'' to identify instances with multiple
unique years, indicating time-series availability. For each such group,
the corresponding ``filtered\_long\_row'' indices are extracted and
stored in a dictionary, facilitating targeted analysis of firms with
repeated observations across years within the same country and industry
classification. This grouped data is then unstacked to pivot the
``country\_official'' and year dimensions, with missing values filled
with zero to ensure a complete matrix. The resulting table is sorted by
``isic'', ``country\_official'', and year for clarity and filtered to
retain only values exceeding 100, with rows and columns containing
exclusively zeros or NaN values removed to focus on significant
observations. The total count of non-NaN values greater than 100 is
computed to quantify the prevalence of substantial firm-level data
points across industries, countries, and years. This refined dataset,
with a recorded count of significant occurrences, serves as a critical
input for subsequent heterogeneity estimation using the zonotope Gini
volume and normalized maximum entropy methods, ensuring the analysis is
grounded in robust and representative data.

\hspace*{2em}The resulting dataset comprises a cross-sectional sample of firms from
economies, covering the years, with a total of firm-level observations
after preprocessing. This cleaned and standardized dataset provides a
robust foundation for estimating TFP using both Cobb-Douglas and
translog production function specifications. In this study, we use to
estimate the trends of industrial heterogeneity in countries shown in Table1. 
\begin{center}
    \includegraphics[width=0.9\textwidth]{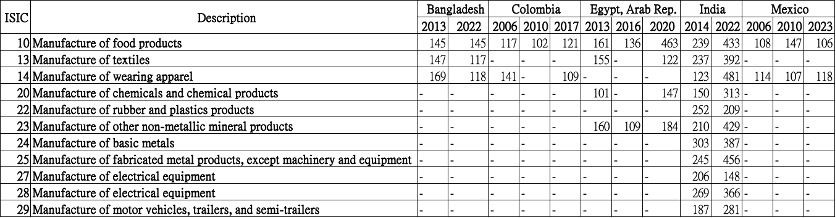}
\end{center}
\sloppy
\hspace*{2em}Table1 provides information on the International Standard Industrial Classification of All Economic Activities (ISIC), a globally recognized framework for classifying economic activities. It includes the 2-digit divisions used in the World Bank dataset, their corresponding industry
descriptions from Section C (Manufacturing), and an overview of the
sample sizes.

\hspace*{2em}In this research, we need information on sales (Y), employment (L), and
capital stock (K). These variables are proxied using the questions
available in the data. More precisely, Y is proxied by total annual
sales of the establishment (variable d2); K is proxied by the
replacement value of machinery, vehicles, and equipment (variable n7a);
and L is proxied by the total annual cost of labor (variable n2a) in
WBES data.
\subsection*{2.1 International Comparability}
\hspace*{2em}All the variables used for the productivity estimation are collected in
local currency units (LCUs), which are specific to the survey and year.
Consequently, the data span different fiscal years. For the estimation
of cross-economy regressions all data must be transformed to a common
currency-year. To do this, all variables are first converted into U.S.
dollars (USD) using the official exchange rate (period average) from the
World Development Indicators (WDI).2 The data are then deflated to 2009
using the GDP deflator for the United States from the relevant reference
fiscal year.3 Note that information on the closing month of the firms'
fiscal year is used to adjust exchange rates and deflators for each
firm.
\subsection*{2.2 Treatment of Outliers}
\hspace*{2em}In this study, we follow World Bank's methods to deal with outliers. To
minimize sensitivity to extreme values, outlier firms are eliminated
from the analysis. More specifically, outliers in d2 (capturing Y), n7a
(capturing K), and n2a (capturing L) were turned into missing before
estimating the production function. To find outliers in levels, we first
transform variables as ln(x+1), and group observations by economy and
broadly defined sector (more precisely, manufacturing and services).
Next, we calculate (unweighted) means and standard deviations of these
transformed variables within each group. Observations that are more than
three standard deviations away from the mean are then marked as outliers
and turned into missing. To find outliers in ratios, we first transform
variables as ln(x), and group observations by industry. The
three-standard-deviation rule is then applied (unweighted) and the
corresponding observations are turned into missing.
\subsection*{2.3 Item Nonresponse}
\hspace*{2em}According to World Bank, another challenge in estimating total factor
productivity is dealing with item nonresponse: i.e., sampled firms do
not answer specific questions of the survey. For example, respondents
may answer the employment question but not sales. To reduce item
non-response, the Enterprise Surveys team follows a strict quality
control process that identifies non-responses and contacts firms to
complete the data. Despite this effort, like many other firm-level
datasets, the ES also suffers from item nonresponse in variables needed
to calculate TFPR. One way to handle the item nonresponse is through
imputation. For example, in the U.S. Census Bureau's 2007 manufacturing
data 73\% of observations have imputed data for at least one variable
used to compute TFPR (White et al. 2018). While item non-response may be
consequential for most analysis, we do not attempt to address it in the
data. We do not employ any of the available imputation or re-weighting
methods that assume that data ``missingness'' is not ignorable and is
related to underlying firm characteristics (Little \& Rubin 2019).
Additionally, the survey (probability) weights included in the data are
agnostic to item non-response and the missingness of productivity
estimates: weights are not re-adjusted or scaled to account for this
missingness.
\subsection*{3. Methods}
\subsection*{3.1 Maximum Entropy}
\hspace*{2em}Entropy measures the uncertainty or randomness in a probability
distribution. A distribution with higher entropy represents greater
uncertainty or lack of information, while lower entropy indicates more
certainty or information. On the other hand, maximum entropy refers to a
principle from information theory and statistical mechanics that is used
to infer the most unbiased probability distribution based on given
constraints. From an economics viewpoint, maximum entropy can be applied
to represent industrial heterogeneity.
For example, if we only know the average firm size in an industry, the
maximum entropy distribution would spread the probability across various
firm sizes as evenly as possible, given this constraint. This leads to
the least biased representation of the industrial landscape, capturing
the uncertainty and heterogeneity effectively. Imagine an industry where
the total revenue is known, and the average firm revenue is given.
Without detailed data on every firm's revenue, maximum
entropy can be used to infer the most probable distribution of revenues
across firms. This might predict that a few firms control a significant
portion of the market, while most firms earn below the average---a
common feature in many real-world industries.
\subsection*{3.1.1 ME regression functions}
\hspace*{2em}A regression function for two inputs $K$ and $L$ where $K$ represents the
replacement value of machinery, vehicles, and equipment and $L$ represents
the total annual cost of labor is defined as
\[
y(K,L) = \int y f(y \mid K, L) \, dy = \frac{\int y f(K,L,y) \, dy}{f(K,L)}
\]
and maximum entropy can be defined as
\[
E_{\max} = - \int y(K,L) \log y(K,L) \, d\mu(K,L)
\]
satisfying
\begin{align}
\int \phi_{\scriptstyle m_k}(K) \phi_{\scriptstyle m_l}(L) y(K,L) \, d\mu(K,L) &= \nu_{\scriptstyle m_k m_l}, \\
d\mu &= f(K,L) \, dK dL
\end{align}
where \[
m_k = 0, \ldots, N_k, \quad
m_l = 0, \ldots, N_l, \quad
\nu_{m_k m_l} = \int \phi_{m_k}(K)\, \phi_{m_l}(L)\, y(K,L)\, d\mu(K,L)
\] 
\[
y(K,L) = \exp\left[
\sum_{n_k=0}^{N_k} \sum_{n_l=0}^{N_l} S_{n_k,n_l} \, \phi_{n_k}(K) \phi_{n_l}(L)
\right]
\]
is estimated
from the observations since
\[
\left( \frac{1}{T} \right)\sum_{t = 1}^{T}{\phi_{m_{k}}\left( K_{t} \right)\phi_{m_{l}}\left( L_{t} \right)}y_{t}
\]
converges to
\(\nu_{m_{k}m_{l}} = \int_{}^{}{\phi_{m_{k}}(K)\phi_{m_{l}}(L)y(K,\ L)d\mu(K,\ L)}\)
as \(T \rightarrow \infty\).
and the solution from the Lagrangian method:
\[
y(K,L) = \exp\left[ \sum_{n_k,n_l=0}^{N_k,N_l} S_{n_k,n_l} \phi_{n_k}(K) \phi_{n_l}(L) \right]
\]
and we approximate \(f(K,\ L)\) with
\[
f(K,L) = \exp\left[ \sum_{n_k,n_l=0}^{N_k,N_l} C_{n_k,n_l} \phi_{n_k}(K) \phi_{n_l}(L) \right]
\]
and then
\[
\nu_{m_k m_l} = \int \phi_{m_k}(K) \phi_{m_l}(L) y(K,L) f(K,L) dK dL
\]
\[
\nu_{m_k m_l} =
\int \phi_{m_k}(K)\, \phi_{m_l}(L)\,
\exp \Biggl[
  \sum_{n_k=0}^{N_k} \sum_{n_l=0}^{N_l} 
    (S_{n_k,n_l} + C_{n_k,n_l}) \, \phi_{n_k}(K) \, \phi_{n_l}(L)
\Biggr]
\, dK \, dL
\]
\subsection*{3.1.2 Computation of Normalized Maximum Entropy}
\hspace*{2em}To estimate industrial heterogeneity using the normalized maximum
entropy (ME) method, we compute the \(H_{max,t}\)metric based on the
distribution of log-transformed output log(Y) across clusters of firms
defined by their input allocations. The process begins by aggregating
log-transformed capital log(K) and labor log(L) into a two-dimensional
input matrix (X). We apply the K-means clustering algorithm with n=10
clusters and a fixed random seed for reproducibility to partition the
firms into groups based on input similarities.
For each cluster, the maximum entropy is approximated by considering the
range and central tendency of log(Y). Specifically, the entropy for
cluster (c) is calculated as:
\[
H_{\mathrm{max},\,\mathrm{cluster},\,c} 
= \log\bigl( y_{\mathrm{max},c} - y_{\mathrm{min},c} \bigr) 
+ \log\bigl( \mathrm{mean}(y_c) \bigr)
\]
where \(y_{max,\ c}\) and \(y_{min,\ c}\) are the maximum and minimum
values of log(Y) within the cluster and mean (\(y_{c}\)) is the mean of
log(Y) in the cluster. An entropy value of zero is assigned if the range
is zero or invalid, ensuring numerical stability. Each cluster is
weighted by its proportion of firms relative to the total sample size,
and the overall normalized maximum entropy \(H_{max,t}\) is obtained as
the weighted average:
\[
H_{\mathrm{max},\,t} = \sum_{c=1}^{n} W_c \, H_{\mathrm{max},\,\mathrm{cluster},\,c},
\quad
W_c = \frac{n_c}{n},
\]
where \(n_c\) is the number of firms in cluster \(c\), and \(n\) is the total number of firms. This metric captures the dispersion and central
tendency of log(Y) across clusters, reflecting the heterogeneity in
production structures influenced by input allocations. The
implementation is executed using Python, leveraging the scikit-learn
library for K-means clustering, and the results are reported for each
dataset to facilitate comparison with the zonotope Gini volume method.
\subsection*{3.1.3 Design of experiment with numerical simulations (with
Python)}
\hspace*{2em}To confirm ME regression functions are appropriate production functions,
we refer to Hang K. Ryu's experiment (1993) using CES (Constant
Elasticity of Substitution) production function as baseline and conduct
the numerical simulations with Python to see R square performance. If
the value of R square is close to 1, which means the approximations of
ME regression functions are almost similar as the approximations
produced by CES production function. In our result, the R square for
comparing the performance between ME regression functions and CES
production function is about 0.99 and the R square between ME regression
functions with polar coordinate representation and CES production
function is around 0.98. Thus, we confirm that ME regression functions
are qualified production functions to estimate approximations and then
we can have confidence to calculate maximum entropy by using the ME
regression functions to evaluate industrial heterogeneity. (Python code
is provided as appendix in the end.)
In numerical simulations for sample size 2000, input values of capital
(\(K_{t})\) and labor (\(L_{t})\ \)were generated by Python NumPy's
random function (np,random.rand) with a domain of {[}0, 1{]} and output
\(y_{t}\) is generated from CES production function.
\(y_{t}\) (from ES production function):
\[
y_t = \gamma \Bigl[ \delta K_t^{-\rho} + (1-\delta) L_t^{-\rho} \Bigr]^{- \upsilon / \rho} e^{u_t},
\]
where the parameters are set as
\[
\gamma = \exp[1.0564], \quad
\delta = 0.4064, \quad
\upsilon = 0.8222, \quad
\rho = 0.6042,
\]
following Kmenta (1986) as cited in Hang K. Ryu's paper, these parameter values came from Kmenta (1986). The residual
\(u_t\) was also generated by Python NumPy's random function
(np.random.rand) satisfying 10 \(u_t \sim \mathcal{N}(0, 1)\)
following a normal distribution.
The R square is defined as:
\[
R^2 \equiv 1 - \frac{\sum_{t=1}^{T} (y_t - \widehat{y}_t)^2}{\sum_{t=1}^{T} y_t^2 - T \, \overline{y}^2},
\]
where \[
\overline{y} \equiv \frac{1}{T} \sum_{t=1}^{T} y_t
\] is the sample mean of \(y_t\), and \(\widehat{y}_t\) is from either ME regression functions or ME regression functions with polar coordinate representation.
A higher entropy value signifies a more heterogeneous industry with
greater competition and diversity among firms, while a lower entropy
value reflects higher concentration and less diversity.
In summary, Maximum entropy reflects heterogeneity in firm behavior,
strategies, and products. An industry with high entropy implies that
firms are offering diverse products, adopting varied strategies, and
possibly serving different market segments. This diversity can be
important for understanding competitive dynamics and innovation within
the industry.
\subsection*{3.2 Zonotopes}
\hspace*{2em}A zonotope is a special type of convex polytope in geometry, defined as
the Minkowski sum of line segments. More intuitively, it can be
understood as a geometric object that results from translating and
combining several line segments (or vectors) in space. The Minkowski sum
of two sets $A$ and $B$ in a vector space is defined as:
\[
A + B = \left\{ a + b \;\middle|\; a \in A,\, b \in B \right\}
\]
\hspace*{2em}A zonotope can be represented as the sum of a finite number of line segments. Suppose 
$v_1, v_2, \dots, v_k$ are vectors in $n$-dimensional space $\mathbb{R}^{n}$. Then the 
zonotope $Z$ generated by these vectors is the set of all points of the form:
\[
Z = \left\{ \sum_{i=1}^{k} \lambda_i v_i \;\middle|\; \lambda_i \in [0,1] \right\},
\]
where $\lambda_i$ are scalar coefficients, typically constrained to lie between 0 and 1. 
This means that the zonotope is the convex hull of all possible combinations of these 
vectors scaled by coefficients between 0 and 1.
In Giovanni Dosi, Marco Grazzi, Luigi Marengo, and Simona Settepanella (2016), they proposed that the (normalized) volume of a zonotope composed by 
vectors representing firms can be used as an indicator of industrial 
heterogeneity. They noted that Hildenbrand (1981) introduces a short-run 
feasible industry production function defined by means of a zonotope 
generated by the family $\{a_n\}_{1 \le n \le N}$ of production activities. 
Let $\{a_n\}_{1 \le n \le N}$ be a collection of vectors in 
$\mathbb{R}^{l+1},\ N \ge l+1$. To any vector $a_n$, a line segment can 
be defined as:
\[
[0, a_n] = \left\{ \chi_n a_n \;\middle|\; \chi_n \in \mathbb{R},\ 0 \le \chi_n \le 1 \right\}.
\]
\hspace*{2em}The short-run total production set associated with the family 
$\{a_n\}_{1 \le n \le N}$ is the Minkowski sum 
$Y = \sum_{n=1}^{N} [0, a_n]$ of line segments generated by production 
activities $\{a_n\}_{1 \le n \le N}$, and the zonotope can be defined as:
\[
Y = \left\{ y \in \mathbb{R}_+^{l+1} \;\middle|\; y = \sum_{n=1}^{N} \phi_n a_n, \ 0 \le \phi_n \le 1 \right\}.
\]
\subsection*{3.2.1 Volume of Zonotopes and Heterogeneity}  
\hspace*{2em}Let $A_{i_l, \dots, i_{l+1}}$ be the matrix whose rows are vectors 
$\{a_{i}, \dots, a_{i_{l+1}}\}$, and let 
$\Delta_{i_l, \dots, i_{l+1}}$ be its determinant. Then, the volume of 
the zonotope $Y$ in $\mathbb{R}^{l+1}$ is defined as:
\[
\mathrm{Vol}(Y) = \sum_{1 \le i_l \le \dots \le i_{l+1} \le N} 
\left| \Delta_{i_l, \dots, i_{l+1}} \right|,
\]
where $\left| \Delta_{i_l, \dots, i_{l+1}} \right|$ is the absolute value of
the determinant $\Delta_{i_l, \dots, i_{l+1}}$.  
The larger the volume of the zonotope, $\mathrm{Vol}(Y)$, the more diverse 
the industry is.
\subsection*{3.2.2 Gini Volume and Heterogeneity}  
\hspace*{2em}The Gini index (or Gini coefficient) is a statistical measure of 
inequality within a distribution, where 0 indicates perfect equality 
and 1 indicates perfect inequality. Dosi et al. (2016) introduced a 
generalization of the Gini index, called the \emph{Gini volume} of a 
zonotope.  
Specifically, consider the zonotope $Y$ generated by the production 
activities $\{a_n\}_{1 \le n \le N}$. Let $d_Y = \sum_{n=1}^{N} a_n$ 
denote the diagonal of the industry’s production activities, and let 
$P_Y$ be the parallelotope with diagonal $d_Y$, whose volume 
$Vol(P_Y)$ represents the maximum volume achievable in the industry. 
The Gini volume is then defined as:
\[
G(Y) = \frac{Vol(Y)}{Vol(P_Y)}.
\]
\hspace*{2em}The closer the Gini volume $G(Y)$ is to 1, the more diverse the 
industry is. Zonotopes provide a geometric representation of diversity 
in industrial heterogeneity: a large volume indicates high heterogeneity, 
while a small volume indicates low heterogeneity.
\subsection*{3.2.3 Tangent Against Input Axes and Technology Change}
\hspace*{2em}To assess the direction and magnitude of technical change within
industries characterized by firm-level heterogeneity, we employ the
"Tangent Against Input Axes" method, a geometric approach that extends
the zonotope-based framework for analyzing production functions. This
method evaluates the orientation of the main diagonal of the
zonotope---a convex polytope formed by the vectors representing
firms---input-output combinations---relative to the
input axes, providing a dynamic measure of technical change over time.
The approach accommodates heterogeneous technological trajectories
across firms without imposing restrictive assumptions such as constant
returns to scale or perfect input substitutability.
The zonotope is constructed from firm-level data, where each firm 
$(\iota)$ is represented by a vector $Z_i = (x_i, y_i)$ in an 
$n$-dimensional input space 
$x_i = (x_{i1}, x_{i2}, \ldots, x_{in})$ (e.g., capital, labor) 
and an $m$-dimensional output space $y_i$. The zonotope $Z$ is defined 
as the Minkowski sum of these vectors:
\[
Z = \sum_{i=1}^{N} [-Z_i, Z_i],
\]
where $N$ is the number of firms, and the zonotope encapsulates the 
range of feasible input-output combinations. The main diagonal of the 
zonotope, connecting the origin to the vertex representing the sum of 
all $Z_i$, is approximated by the vector 
\[
d = \sum_{i=1}^{N} Z_i.
\]
\hspace*{2em}The direction of technical change is inferred from the angles 
$\theta_j$ (for $j=1,\ldots,n$) between this diagonal and each input 
axis $j$.  
The tangent angle $\theta_j$ with respect to the $j$-th input axis is 
computed as:
\[
\theta_j = \arctan \left( \frac{\| d_{\bot j} \|}{\| d_{\parallel j} \|} \right),
\]
where $d_{\parallel j}$ is the projection of $d$ onto the $j$-th input 
axis, and $d_{\bot j}$ is the component of $d$ perpendicular to that 
axis. These components are derived from the coordinates of $d$:
\[
d_{\parallel j} = (d_j, 0, \ldots, 0), \quad d_{\bot j} = d - d_{\parallel j},
\]
with $d_j = \sum_{i=1}^{N} x_{ij}$ being the sum of the $j$-th input 
across all firms. The magnitudes of the projections are:
\[
\| d_{\parallel j} \| = |d_j|, \quad 
\| d_{\bot j} \| = \sqrt{\sum_{k \neq j} d_k^2 + \sum_{m=1}^{m} d_{y_m}^2},
\]
where $d_{y_m}$ are the output components. The angle $\theta_j$ is 
expressed in radians, with $\theta_j \approx 0$ indicating strong 
reliance on input $j$, and $\theta_j \approx \pi/2$ suggesting reduced 
dependence due to technical progress.  
Changes in $\theta_j$ over time capture shifts in input utilization, 
with a decreasing angle indicating increased efficiency or substitution 
away from input $j$, and an increasing angle suggesting sustained or 
growing dependence. This method complements the static heterogeneity 
index derived from the zonotope volume, offering a dynamic perspective 
on how technical advancements reshape industrial production structures 
across heterogeneous firms.  

\subsection*{4. Results}

\subsection*{4.1 Empirical Findings}
\hspace*{2em}The results are illustrated in the following graphs. Similar trends are 
observed between Gini Volume and our proposed method, Normalized Maximum 
Entropy, in Bangladesh, Colombia, Egypt (Arab Rep.), and India.
\begin{figure}[ht!]
    \centering
    \includegraphics[width=0.9\textwidth]{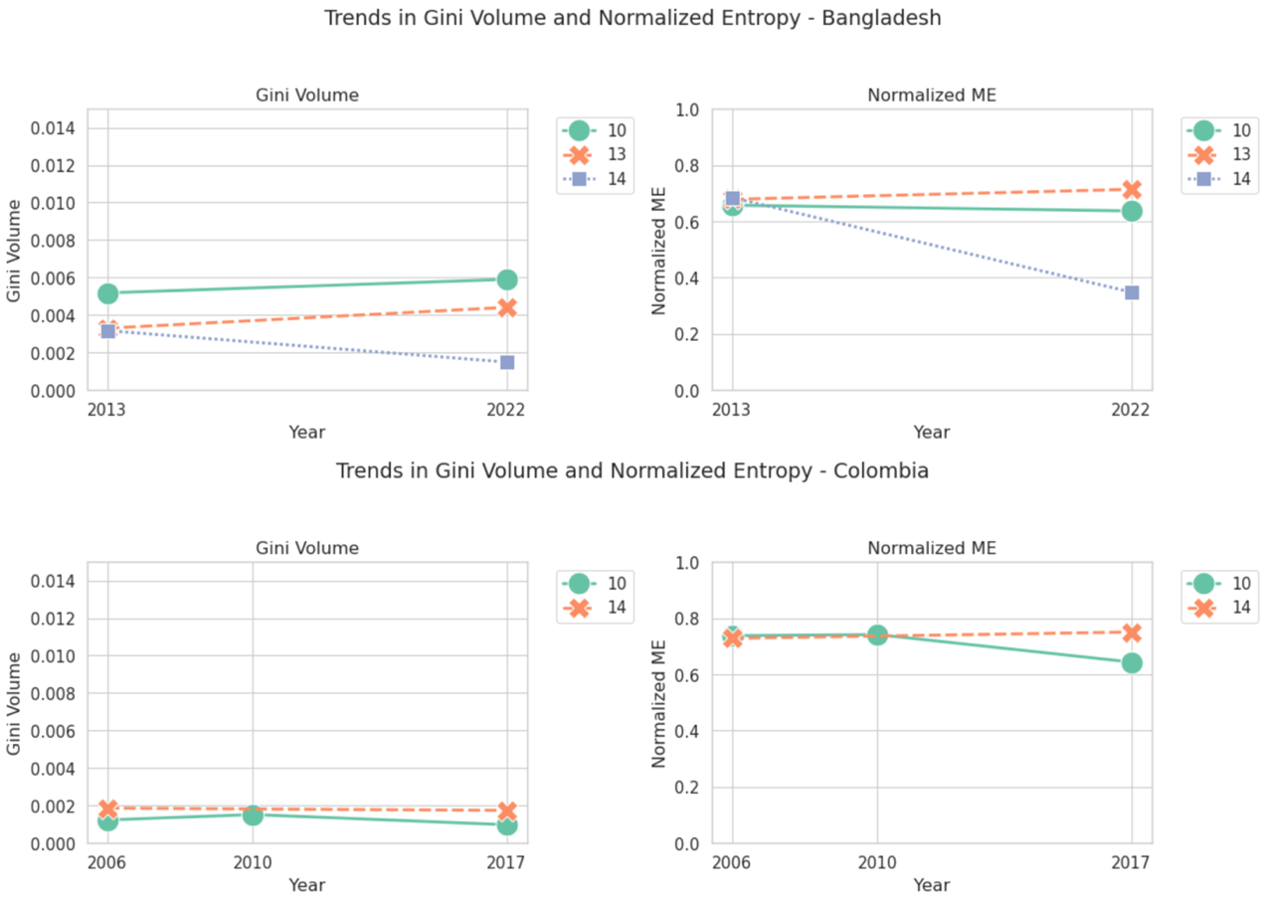}
\end{figure}
\begin{figure}[ht!]
    \centering
    \includegraphics[width=0.9\textwidth]{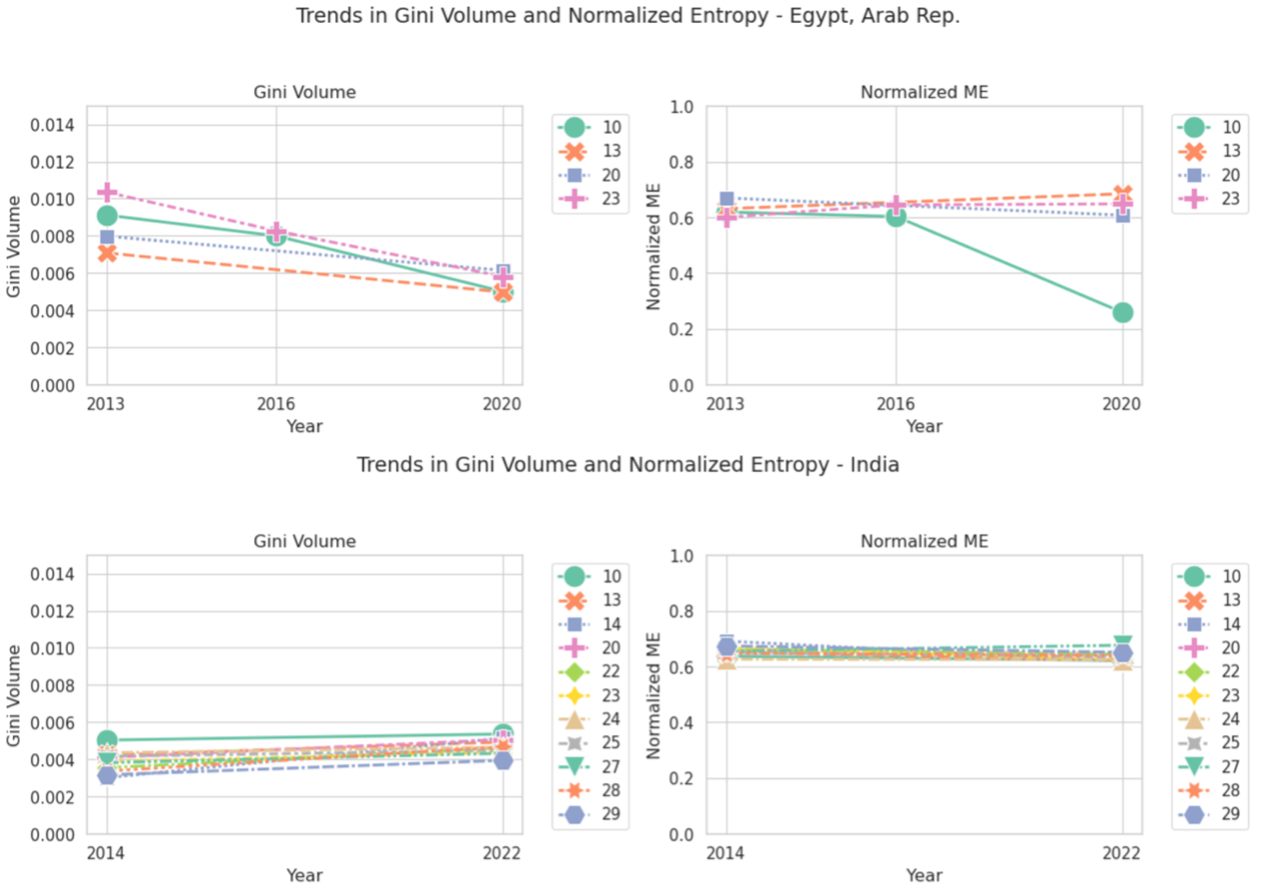}
\end{figure}
\sloppy
\hspace*{2em}Mexico is the only country that exhibits an opposite trend---Gini Volume decreases over time, whereas Normalized Maximum Entropy increases.
Additional analysis is needed to clarify heterogeneity in
Mexico is increasing or decreasing.

\hspace*{2em}To illustrate the contrasting trends observed in the zonotope-based Gini
value and Normalized ME, we construct a stylized example with three
firms in Mexico's ISIC 10 industry. It is crucial to
recognize that this is a simplified representation of complex firm
dynamics. In particular, firm-level labor productivity
('a') is calculated as a residual, which
is subject to potential endogeneity and omitted variable bias. While
this approach allows for a clear demonstration of the differences
between the two heterogeneity measures, the conclusions drawn from this
example should be interpreted with caution. We acknowledge that in
reality, ('a') is influenced by numerous
other factors not explicitly accounted for here, such as technology
adoption, management practices, and access to resources.

\hspace*{2em}This discrepancy may stem from structural changes in
Mexico's economy that are not fully captured by income
distribution alone. For instance, shifts in industrial composition,
regional disparities, or informal labor dynamics could influence
entropy-based measures differently than traditional inequality metrics.
\hspace*{2em}In next section, another indicator, Tangent Against Input Axes, provides
more insight about the nature of heterogeneity in Mexico and to
reconcile the divergent trends observed in Gini Volume and Normalized
Maximum Entropy.
\begin{figure}[ht!]
    \centering
    \includegraphics[width=0.9\textwidth]{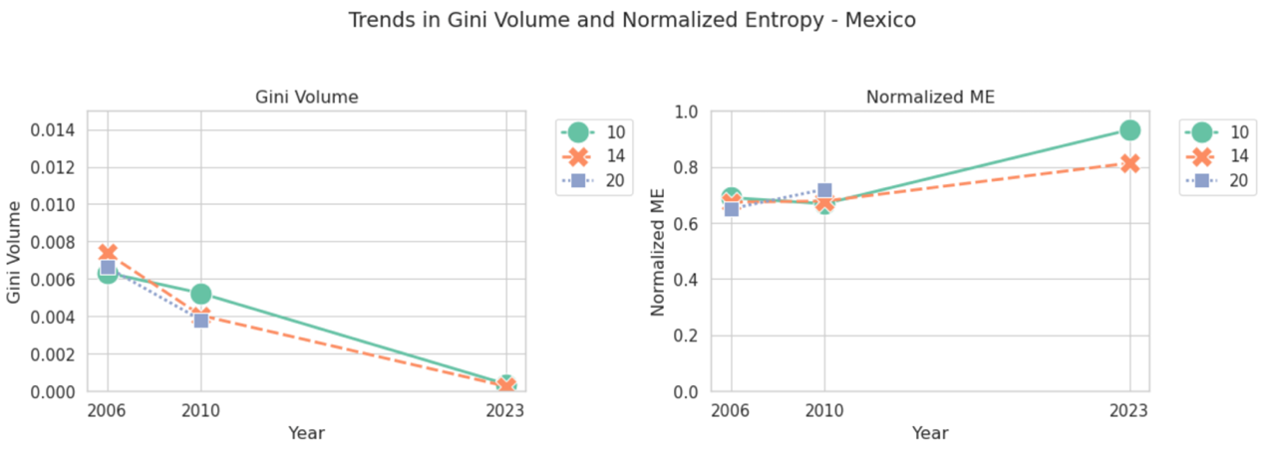}
    \caption{Visualization of Heterogeneity Measures}
    \label{fig:image4}
\end{figure}

\subsection*{4.1.1 Contrasting Results and Analysis}

\hspace*{2em}The results reveal a striking contrast between the two measures:

Table 2: Heterogeneity Measures for Mexico's ISIC 10 Industry

\begin{longtable}{@{}p{1.2cm}p{3.5cm}p{2.5cm}p{3.8cm}@{}} 
\toprule
\textbf{Year} & \textbf{Zonotope-Based Gini Value} & \textbf{Normalized ME} & \textbf{Tangent Against Input Axes} \\
\midrule
\endfirsthead

\multicolumn{4}{c}{{\bfseries Table \thetable\ continued from previous page}} \\ 
\toprule
\textbf{Year} & \textbf{Zonotope-Based Gini Value} & \textbf{Normalized ME} & \textbf{Tangent Against Input Axes} \\
\midrule
\endhead

\bottomrule
\endfoot

\bottomrule
\endlastfoot
2006 & 0.0063 & 0.6996 & 0.8527 \\
2010 & 0.0052 & 0.6677 & 0.8210 \\
2023 & 0.0004 & 0.9320 & 1.0844 \\
\end{longtable}

\hspace*{2em}The zonotope-based Gini value decreases significantly from 0.0063 in
2006 to 0.0004 in 2023, suggesting a substantial reduction in firm
heterogeneity. This indicates that firms in the ISIC 10 industry have
become more homogeneous in terms of their input-output vectors (Y, K,
L), likely due to industry standardization, technological diffusion, or
policy interventions.

\hspace*{2em}In contrast, the Normalized ME increases from 0.6677 in 2010 to 0.9320
in 2023, indicating a rise in functional heterogeneity. This suggests
that, despite the convergence in inputs and outputs, the distribution of
Y=f (K, L) has become more disordered, reflecting greater diversity in
how firms transform inputs into outputs.

\hspace*{2em}Additionally, the tangent against input axes, which measures the rate of
technical change relative to input axes, increases from 0.8210 in 2010
to 1.0844 in 2023, suggesting that technical change may have contributed
to the observed trends in heterogeneity.

\subsection*{Why the Contrasting Results? A Theoretical and Empirical Analysis}
\hspace*{2em}The contrasting trends in heterogeneity can be attributed to the
fundamental differences in what each method measures:
The zonotope-based Gini value captures quantitative heterogeneity by
measuring the inequality in the distribution of (Y, K, L) vectors. A
decline in the Gini value indicates that firms have become more similar
in their input and output quantities, reflecting a convergence in scale
and production levels.

\hspace*{2em}The maximum entropy method, however, captures functional heterogeneity
by focusing on the disorder (entropy) in the distribution of Y=f (K, L).
An increase in Normalized ME suggests that, even as firms converge in
their input-output quantities, their production functions---i.e., the
efficiency with which they transform inputs into outputs---have become
more diverse.

\hspace*{2em}To understand why functional heterogeneity may increase despite a
convergence in (Y, K, L), consider the following: while industry
standardization may lead firms to adopt similar levels of inputs (K, L)
and achieve similar outputs (Y), technical change can introduce
diversity in production efficiency. For instance, some firms may adopt
advanced technologies that enhance their productivity, while others
continue using traditional methods, leading to differences in how inputs
are transformed into outputs.

\subsection*{4.2 Illustrative Example}

\hspace*{2em}To illustrate this phenomenon, we construct a stylized example with
three firms (A, B, C) in Mexico's ISIC 10 industry, focusing on their
labor input (L), capital input (K), and output (Y). We assume a simple
production function of the form \(Y = aLK\), where (a) represents labor
productivity, and (K) is held constant for simplicity.
\begin{itemize}
\item 2006 (Gini Value = 0.0063, Normalized ME = 0.6996):
The following table presents firm-level data from the WBES:
\begin{longtable}{%
>{\raggedright\arraybackslash}p{0.0961\linewidth}%
>{\raggedright\arraybackslash}p{0.1706\linewidth}%
>{\raggedright\arraybackslash}p{0.1812\linewidth}%
>{\raggedright\arraybackslash}p{0.2300\linewidth}%
>{\raggedright\arraybackslash}p{0.3220\linewidth}}
\toprule
\makecell{Firm} & \makecell{L\\(Workers)} & \makecell{K\\(Machines)} & \makecell{Y\\(Output Units)} & \makecell{a\\(Labor Productivity)} \\
\midrule
\endfirsthead
\toprule
\makecell{Firm} & \makecell{L\\(Workers)} & \makecell{K\\(Machines)} & \makecell{Y\\(Output Units)} & \makecell{a\\(Labor Productivity)} \\
\midrule
\endhead
\bottomrule
\endfoot
\bottomrule
\endlastfoot
\textbf{A} & 100 & 10 & 1200 & 12 \\
\textbf{B} & 100 & 10 & 1300 & 13 \\
\textbf{C} & 100 & 10 & 1100 & 11 \\
\end{longtable}
\end{itemize}
\begin{itemize}
\item
  Input-Output Distribution: (L) ranges from 50 to 150, (K) from 5 to
  20, and (Y) from 1000 to 1500 (range = 500, mean = 1233, coefficient
  of variation {[}CV{]} = 0.20, where CV is defined as the standard
  deviation divided by the mean, providing a standardized measure of
  relative dispersion). The (Y, K, L) vectors are relatively dispersed,
  leading to a higher Gini value (0.0063).
\item
  Production Function: Labor productivity (a) ranges from 8 to 30 (range
  = 22, CV = 0.696). The output (Y) is influenced by both (L) and (a),
  resulting in a distribution that is relatively ordered (e.g., closer
  to a normal distribution), with a lower entropy (Normalized ME =
  0.6996).
\end{itemize}
\begin{itemize}
\item
  2023 (Gini Value = 0.0004, Normalized ME = 0.9320):
\end{itemize}
\begin{longtable}{@{}
  >{\centering\arraybackslash}p{0.10\linewidth}
  >{\raggedright\arraybackslash}p{0.17\linewidth}
  >{\raggedright\arraybackslash}p{0.18\linewidth}
  >{\raggedright\arraybackslash}p{0.23\linewidth}
  >{\raggedright\arraybackslash}p{0.32\linewidth}
@{}}
\end{longtable}
\begin{longtable}{@{}p{1.5cm}cccc@{}}
\toprule
\makecell{Firm} & \makecell{L\\(Workers)} & \makecell{K\\(Machines)} & \makecell{Y\\(Output Units)} & \makecell{a\\(Labor Productivity)} \\
\midrule
\endfirsthead
\toprule
\makecell{Firm} & \makecell{L\\(Workers)} & \makecell{K\\(Machines)} & \makecell{Y\\(Output Units)} & \makecell{a\\(Labor Productivity)} \\
\midrule
\endhead
\bottomrule
\endfoot
\bottomrule
\endlastfoot
\textbf{A} & 100 & 10 & 1200 & 12 \\
\textbf{B} & 100 & 10 & 1300 & 13 \\
\textbf{C} & 100 & 10 & 1100 & 11 \\
\end{longtable}
\begin{itemize}
\item
  Input-Output Distribution: (L) and (K) are identical across firms
  (L=100, K=10), and (Y) ranges from 1100 to 1300 (range = 200, mean =
  1200, CV = 0.083). The (Y, K, L) vectors are highly concentrated,
  leading to a near-zero Gini value (0.0004).
\item
  Production Function: Labor productivity (a) ranges from 11 to 13
  (range = 2, CV = 0.083). Since (L) is fixed, the variation in (Y) is
  entirely driven by (a). The output distribution (1100, 1200, 1300) is
  nearly uniform (each value has a probability of 1/3), resulting in a
  higher entropy (Normalized ME = 0.9320).
\end{itemize}
\subsection*{Why Does Disorder Increase Despite a Smaller Range in (a)?}
\hspace*{2em}The increase in Normalized ME, despite the smaller range in (a), can be
explained by the following factors:
\begin{itemize}
\item
  \textbf{Entropy Depends on Distribution Shape, Not Absolute Range:}
  Entropy measures the disorder of a distribution, not the absolute
  range of values. In 2006, the output (Y) (1000, 1200, 1500) is
  influenced by both (L) and (a), resulting in a distribution that may
  be more ordered (e.g., closer to a normal distribution), with a lower
  entropy. In 2023, (L) is fixed, and (Y) (1100, 1200, 1300) is entirely
  determined by (a), forming a nearly uniform distribution (probability
  1/3 for each value). The entropy of a uniform distribution with three
  values is $H = -\sum_{i=1}^3 \left( \frac{1}{3} \ln \frac{1}{3} \right) = \ln 3 \approx 1.099$, which is higher than a
  more ordered distribution, leading to a higher Normalized ME.
\item
  \textbf{Fixed (L) Amplifies the Effect of (a):} In 2006, the variation
  in (L) (50 to 150) partially offsets the large variation in (a) (8 to
  30), smoothing the output distribution. In 2023, with (L) fixed at
  100, the variation in (Y) directly reflects the variation in (a),
  making the relative disorder more pronounced, even though the absolute
  range of (a) is smaller.
\item
  \textbf{Technical Change as a Driver:} The increase in the tangent
  against input axes (from 0.8210 to 1.0844) suggests that technical
  change has played a role in increasing heterogeneity. Even with a
  smaller range in (a), technical change may have introduced diversity
  in production efficiency. For example, Firm B may have adopted
  advanced automation technologies (a=13), while Firm C continues using
  traditional methods (a=11), leading to differences in output despite
  identical inputs.
\end{itemize}
\subsection*{Supporting the Maximum Entropy Method}
\hspace*{2em}The contrasting results highlight the limitations of the zonotope-based
Gini value and the strengths of the maximum entropy method:
\begin{itemize}
\item
  \textbf{Limitation of the Gini Value:} The Gini value captures only
  quantitative heterogeneity, focusing on the inequality in (Y, K, L)
  vectors. Its decline (from 0.0063 to 0.0004) reflects a convergence in
  inputs and outputs, but it fails to capture the functional
  heterogeneity in how firms transform inputs into outputs. In the
  example, the Gini value underestimates heterogeneity in 2023, as it
  does not account for the diversity in production efficiency (a).
\item
  \textbf{Strength of the Maximum Entropy Method:} The maximum entropy
  method directly estimates the distribution of Y=f (K, L), capturing
  the disorder in the output distribution. The increase in Normalized ME
  (from 0.6677 to 0.9320) accurately reflects the growing functional
  heterogeneity in production efficiency, even as the absolute range of
  (a) decreases. This method is particularly effective in modern
  economic contexts, where heterogeneity often arises from qualitative
  differences (e.g., technological adoption, management practices)
  rather than quantitative disparities in inputs and outputs.
\item
  \textbf{Flexibility and Sensitivity:} The maximum entropy method is a
  non-parametric approach that does not rely on specific distributional
  assumptions (unlike the Gini value, which assumes inequality can be
  represented by a Lorenz curve). It is highly sensitive to changes in
  the shape of the output distribution, making it well-suited to capture
  the nuanced effects of technical change and efficiency differences.
\end{itemize}
  
\subsection*{4.3  Zonotope Issues}

\hspace*{2em}The zonotope method, as outlined in the referenced study, employs the
normalized volume of a zonotope, defined as
\(\frac{Vol(Z_{t})}{Vol(P_{y,t})}\), to quantify industry heterogeneity.
However, our analysis reveals two critical limitations that undermine
its ability to accurately reflect firm-level heterogeneity and track its
temporal trends: (1) bias in the normalization denominator
\(Vol(P_{y,t})\), which leads to underestimation of heterogeneity, and
(2) the dynamic nature of industry entry thresholds, which renders the
metric unsuitable for longitudinal trend analysis. In contrast, the
proposed Maximum Entropy (ME) approach, utilizing normalized entropy
\(H_{norm,\ t} = \frac{H_{t}^{*}}{H_{max,t}}\), offers a more robust and
precise measure of industry heterogeneity. Below, we provide detailed
mathematical derivations to substantiate these findings.

\subsection*{4.3.1 Bias in the Normalization Denominator}

\hspace*{2em}The zonotope \(Z_{t}\) is constructed as the convex hull of
firm-specific input-output vectors \(\upsilon_{i} = (x_{i},\ y_{i})\),
where \(x_{i} \in R^{k}\) represents inputs (e.g., labor, capital) and
\(y_{i} \in R^{m}\) denotes outputs, with:
\[Z_{i} = \left\{ \sum_{i = 1}^{n}{\lambda_{i}\upsilon_{i}\ |}\lambda_{i} \in \lbrack 0,\ 1\rbrack,\ \sum_{i = 1}^{n}{\lambda_{i} = 1} \right\}\]
\hspace*{2em}The volume \(Vol(Z_{t})\) captures the geometric dispersion of these
vectors, serving as a proxy for heterogeneity. The normalized volume is
computed as:
\[{Normalized\ Volume}_{t} = \frac{Vol(Z_{t})}{Vol(P_{y,t})}\]
where \(P_{y,t}\) is the reference output space, typically defined as
the convex hull of observed outputs
\(P_{y,t} = conv\left\{ y_{i} \right\}\) or a theoretical range
\(P_{y,t} = {\lbrack 0,\ max\left\{ y_{i} \right\}\rbrack}^{m}\).
However, if \(P_{y,t}\) includes infeasible outputs---such as
\(y_{i} = 0\) or values below the industry's minimum entry threshold
\(y_{min,t}\)---the denominator \(Vol(P_{y,t})\) becomes inflated,
leading to a biased heterogeneity measure.
Consider a single-output case (m=1) for simplicity. Let the observed
output range be \(y_{i} \in \lbrack y_{min,t},\ y_{max,t}\rbrack\),
where \(y_{min,t} > 0\) reflects the industry's entry threshold (e.g.,
minimum production required to sustain operations). If \(P_{y,t}\) is
defined as \(\lbrack 0,\ y_{max,t}\rbrack\), the volume of the reference
space is:
\[Vol\left( P_{y,t} \right) = y_{max,t} - 0 = y_{max,t}\]
\hspace*{2em}However, the realistic range, accounting for the entry threshold, is
\(\lbrack y_{min,t},\ y_{max,t}\rbrack\), with volume:
\[Vol\left( P_{y,t}^{*} \right) = y_{\max,t} - y_{\min,t}\]
Since \(y_{\min,t} > 0\), it follows that:
\[Vol\left( P_{y,t} \right) > \ Vol\left( P_{y,t}^{*} \right)\]
\hspace*{2em}Consequently, the normalized volume is underestimated:
\[\frac{Vol\left( Z_{t} \right)}{Vol\left( P_{y,t} \right)} < \frac{Vol\left( Z_{t} \right)}{Vol\left( P_{y,t}^{*} \right)}\]
\hspace*{2em}For example, if \(y_{\min,t} = 100,\ y_{\max,t} = 1000\), and
\(Vol\left( Z_{t} \right) = 500\), the biased normalized volume is:
\[\frac{500}{1000} = 0.5\]
while the true measure, using the threshold-adjusted range, is:
\[\frac{500}{1000 - 100} = \frac{500}{900} \approx 0.556\]
\hspace*{2em}This underestimation distorts the heterogeneity measure, as the
inclusion of infeasible outputs (\(y_{i} < y_{\min,t}\)) artificially
inflates the denominator, suppressing the perceived diversity of firm
production profiles.

\subsection*{4.3.2 Unsuitability for Temporal Trend Analysis}

\hspace*{2em}The second limitation arises from the annual variation in industry entry
thresholds, which introduces dynamic biases in
\(Vol\left( P_{y,t} \right)\), rendering the normalized volume
unsuitable for tracking heterogeneity trends over time. Industry entry
thresholds denoted \(y_{\min,t}\), evolve due to technological
advancements, market conditions, or regulatory changes. For instance, in
a manufacturing sector, \(y_{\min,t}\) may increase over time as
automation raises the minimum capital requirement. If \(P_{y,t}\) is
defined without accounting for these shifts (e.g., consistently assuming
\(y_{i} \in \lbrack 0,\ \ max\left\{ y_{i} \right\}\rbrack\)), the bias
in \(Vol\left( P_{y,t} \right)\) varies across years, compromising
cross-year comparability. Mathematically, consider two years \(t_{1}\)
and \(t_{2}\), with entry thresholds \(y_{\min,t_{1}} < y_{\min,t_{2}}\)
and maximum outputs \(y_{\max,t_{1}} \approx y_{\max,t_{2}}\). If
\(P_{y,t} = \lbrack 0,\ y_{\max,t}\rbrack\), then:
\[Vol\left( P_{y,t_{1}} \right) = y_{\max,t_{1}},\ \ Vol\left( P_{y,t_{2}} \right) = y_{\max,t_{2}}\]
\hspace*{2em}However, the realistic volumes are:
\[Vol\left( P_{y,t_{1}}^{*} \right) = y_{\max,t_{1}} - y_{\min,t_{1}},\ \ Vol\left( P_{y,t_{2}}^{*} \right) = y_{\max,t_{2}} - y_{\min,t_{2}}\ \]
Since \(y_{\min,t_{2}} > y_{\min,t_{1}}\), the bias in \(t_{2}\) is more
pronounced:
\[\frac{Vol\left( P_{y,t_{2}} \right)}{Vol\left( P_{y,t_{2}}^{*} \right)} < \frac{Vol\left( P_{y,t_{1}} \right)}{Vol\left( P_{y,t_{1}}^{*} \right)}\]
\hspace*{2em}This results in a greater underestimation of the normalized volume in
\(t_{2}\):
\[\frac{Vol\left( Z_{t_{2}} \right)}{Vol\left( P_{y,t_{2}} \right)} < \frac{Vol\left( Z_{t_{2}} \right)}{Vol\left( P_{y,t_{2}}^{*} \right)}\ and\ \frac{Vol\left( Z_{t_{2}} \right)}{Vol\left( P_{y,t_{2}} \right)} < \frac{Vol\left( Z_{t_{1}} \right)}{Vol\left( P_{y,t_{1}} \right)}\ \]
even if the true heterogeneity (reflected by
\(\frac{\mathrm{Vol}\left( Z_{t} \right)}{\mathrm{Vol}\left( P_{y,t}^{*} \right)}\))
remains constant or increases. 
\hspace*{2em}Consequently, the observed trend may
erroneously suggest a decline in heterogeneity, driven by the varying
bias in \(\mathrm{Vol}\left( P_{y,t} \right)\) rather than actual changes in firm
diversity.

\subsection*{4.3.3 Superiority of the Maximum Entropy Approach}

\hspace*{2em}In contrast, the proposed Maximum Entropy (ME) method offers a robust
alternative for measuring industry heterogeneity, leveraging the
normalized entropy
\[
H_{\mathrm{norm}, t} = \frac{H_{t}^{*}}{H_{\max,t}}.
\]
\hspace*{2em}The ME approach estimates a nonparametric production function
\[
P\left( y \,\middle|\, x,t \right)
\]
by maximizing the entropy:
\[
H_{t}^{*} = - \int P\left( y \,\middle|\, x,t \right) \log P\left( y \,\middle|\, x,t \right) \, dy,
\]
subject to moment constraints derived from the data. The maximum
possible entropy \(H_{\max,t}\) is calculated based on the conditional
output range
\([y_{\min}(x,t),\ y_{\max}(x,t)]\),
which is determined empirically for each input vector \(x\). For a continuous
output \(y\), the maximum entropy under a uniform distribution is:
\[
H_{\max,t}\left( Y \,\middle|\, x,t \right) = \log\Big( y_{\max}(x,t) - y_{\min}(x,t) \Big).
\]
and the total maximum entropy is the expectation over the input
distribution:
\[
H_{\max,t} = \int P(x,t) \, \log \Big( y_{\max}(x,t) - y_{\min}(x,t) \Big) \, dx.
\]
\hspace*{2em}In practice, this is approximated by discretizing \(x\) into clusters
(e.g., via K-means) and computing:
\[
H_{\max,t} \approx \sum_{k=1}^{K} P(x_k,t) \, \log \Big( y_{\max}(x_k,t) - y_{\min}(x_k,t) \Big),
\]
where \(P(x_k,t)\) is the sample proportion in cluster \(k\).
Unlike \(Vol(P_{y,t})\), which may include infeasible outputs,
\(H_{\max,t}\) is derived from observed data, ensuring that
\(y_{\min}(x,t)\) reflects the industry's entry threshold. For instance,
if the minimum output threshold increases from \(y_{\min, t_1} = 100\)
to \(y_{\min, t_2} = 150\), the ME method recalibrates \(H_{\max,t}\)
based on the updated ranges, avoiding the inclusion of unrealistic values.
This data-driven approach mitigates the bias seen in the zonotope method,
as demonstrated by:
\[
H_{\mathrm{norm}, t} = \frac{H_t^*}{\sum_{k=1}^{K} P(x_k,t) \, \log \Big( y_{\max}(x_k,t) - y_{\min}(x_k,t) \Big)},
\]
which remains bounded in \([0,1]\) and is comparable across years,
provided data structures are consistent.
Moreover, the ME method adapts to annual threshold variations by
recalculating \(H_{\max,t}\) each year, ensuring that temporal
comparisons reflect true changes in heterogeneity rather than artifacts
of normalization bias. Preliminary results from applying the ME method
to [insert industry or dataset, e.g., manufacturing sector data from
2015--2020] indicate that \(H_{\mathrm{norm},t}\) effectively captures the
evolution of production uncertainty, aligning with economic indicators
such as technological adoption or market concentration. In contrast, the
zonotope method's normalized volume exhibited inconsistent trends,
underestimating heterogeneity in years with higher entry thresholds due
to the inflated \(Vol\left( P_{y,t} \right)\). These findings underscore
the ME method's superior ability to provide a precise and economically
meaningful measure of industry heterogeneity, particularly for
longitudinal analyses.

\subsection*{5. Simulation}

\hspace*{2em}From our empirical findings, we observed that the zonotope method may
have certain limitations. To investigate this further, we designed an
experiment involving two datasets: one with high heterogeneity,
generated from a uniform distribution, and another with low
heterogeneity, generated from a lognormal distribution. To assess the
effectiveness of the zonotope Gini volume and normalized maximum entropy
(ME) methods in estimating industrial heterogeneity, we conducted a
Monte Carlo simulation using synthetic data derived from a Cobb-Douglas
production function. The simulation provides evidence supporting that
the Gini volume method tends to be biased---especially in capturing the
underlying randomness of production structures---whereas the ME method
offers a more robust measure of heterogeneity.

\subsection*{5.1 Data Generation}

\hspace*{2em}We generate 100 pairs of datasets, each comprising n =100 firms,
representing low and high heterogeneity in industrial production. The
Cobb-Douglas production function is specified as:
\[Y_{i} = A_{i}K_{i}^{\alpha}L_{i}^{\beta},\]
where \(Y_{i}\) is the output of firm (i), \(K_{i}\) and \(L_{i}\) are
capital and labor inputs, \(A_{i}\) is total factor productivity, and
the elasticities are fixed at $\alpha = 0.33$ and $\beta = 0.66$. A fixed random seed ensures reproducibility.
\begin{itemize}
\item
\textbf{High-Heterogeneity Dataset}: To simulate high heterogeneity, we impose a structured and uniform allocation of inputs to reflect a production environment with minimal randomness in input variation. Capital and labor are drawn from:
\begin{align}
K_{i} &\sim \text{Uniform}(2900,3100), \\
L_{i} &\sim \text{Uniform}(120,130),
\end{align}
ensuring tight ranges with a variation of approximately 6.9\% and 8.3\% around their means, respectively. Total factor productivity is sampled from:
\begin{equation}
A_{i} \sim \text{Uniform}(1.5,2.5).
\end{equation}
This configuration introduces moderate and consistent productivity variation, yielding relatively uniform outputs \( Y_{i} \).
  reflecting a low-dispersion production structure.
\item
\textbf{Low-Heterogeneity Dataset}: To simulate low heterogeneity, we
  introduce a structured yet dispersed input allocation, characterized
  by lognormal distributions that impose a predictable skewed pattern.
  Capital \(K_{i}\ \)is drawn from a lognormal distribution with mean
  \(\mu_{K} = \log(3000) - {0.5(0.01344)}^{2}\) and standard deviation
  \(\sigma_{K} = 0.01344\), while labor \(L_{i}\ \)follows a lognormal
  distribution with mean \(\mu_{L} = \log(125) - {0.5(0.01661)}^{2}\)
  and standard deviation \(\sigma_{L} = 0.01661\). The standard
  deviations \(\sigma_{K} = 0.01344\) and \(\sigma_{L} = 0.01661\ \)are
  calibrated to ensure the means of \(K_{i}\) and \(L_{i}\) approximate
  3000 and 125, respectively, after accounting for the lognormal
  adjustment, where the mean of a lognormal variable is
  \(exp(\mu + \frac{\sigma^{2}}{2})\). Solving for \(\mu_{K}\) with
  \(\sigma_{K} = 0.01344\), we obtain \(\mu_{K} \approx 8.0059\) (where
  \( \log(3000) \approx 8.006 \), and for \( \sigma_{L} = 0.01661 \),
  \( \mu_{L} = 4.8279 \) (where \( \log(125) \approx 4.828 \)), ensuring the expected
  values align with the target means while introducing controlled
  skewness. We generate 100 pairs of datasets, each comprising \(n = 100\) firms,
representing low and high heterogeneity in industrial production. The
Cobb-Douglas production function is specified as:
\[
Y_{i} = A_{i} K_{i}^{\alpha} L_{i}^{\beta},
\]
where \(Y_{i}\) is the output of firm \(i\), \(K_{i}\) and \(L_{i}\) are
capital and labor inputs, \(A_{i}\) is total factor productivity, and
the elasticities are fixed at \(\alpha = 0.33\) and \(\beta = 0.66\).
A fixed random seed ensures reproducibility.
\end{itemize}
\begin{itemize}
\item
  \textbf{High-Heterogeneity Dataset}: To simulate high heterogeneity,
  we impose a structured and uniform allocation of inputs to reflect a
  production environment with minimal randomness in input variation.
  Capital \(K_{i}\) and labor \(L_{i}\) are drawn from uniform
  distributions: \(K_{i} \sim \text{Uniform}(2900,3100)\) and \(L_{i} \sim \text{Uniform}(120,130)\),
  ensuring tight ranges with a variation of approximately 6.9\% and 8.3\% 
  around their means, respectively. Total factor productivity is sampled 
  from a uniform distribution \(A_{i} \sim \text{Uniform}(1.5,2.5)\), 
  introducing moderate and consistent productivity variation. This configuration 
  yields relatively uniform outputs \(Y_{i}\), reflecting a low-dispersion 
  production structure.
\item
  \textbf{Low-Heterogeneity Dataset}: To simulate low heterogeneity, we
  introduce a structured yet dispersed input allocation, characterized
  by lognormal distributions that impose a predictable skewed pattern.
  Capital \(K_{i}\) is drawn from a lognormal distribution with mean
  \(\mu_{K} = \log(3000) - \frac{1}{2} (0.01344)^{2}\) and standard deviation
  \(\sigma_{K} = 0.01344\), while labor \(L_{i}\) follows a lognormal
  distribution with mean \(\mu_{L} = \log(125) - \frac{1}{2} (0.01661)^{2}\)
  and standard deviation \(\sigma_{L} = 0.01661\). These standard deviations
  are calibrated to ensure the means of \(K_{i}\) and \(L_{i}\) approximate
  3000 and 125, respectively, after accounting for the lognormal adjustment,
  where the mean of a lognormal variable is \(\exp(\mu + \frac{\sigma^{2}}{2})\). 
  Solving for \(\mu_{K}\) with \(\sigma_{K} = 0.01344\), we obtain \(\mu_{K} \approx 8.0059\) 
  (since \(\log(3000) \approx 8.006\)), and for \(\sigma_{L} = 0.01661\), 
  \(\mu_{L} \approx 4.8279\) (since \(\log(125) \approx 4.828\)), ensuring 
  the expected values align with the target means while introducing controlled skewness. 
  The small values of \(\sigma_{K}\) and \(\sigma_{L}\) result in a 95\% range 
  of approximately 2980 to 3020 for \(K\) and 123 to 127 for \(L\), producing 
  a moderately dispersed yet skewed output \(Y_{i}\) when combined with total 
  factor productivity generated as \(A_{i} = \exp(\widetilde{A_{i}})\), where
  \(\widetilde{A_{i}} \sim N(0,0.5)\), introducing moderate variation. This setup 
  reflects greater dispersion and a structured heterogeneity in production processes.
\end{itemize}
\hspace*{2em}Each dataset, containing \(K_{i}\), \(L_{i}\), and \(Y_{i}\), is saved
as i-high\_heterogeneity\_data.csv and i-low\_heterogeneity\_data.csv
(for i=1\ldots,100) in the directory data-100 for subsequent analysis.


\subsection*{5.2 Simulation Design}
\hspace*{2em}For each pair of datasets, we estimate industrial heterogeneity using
the zonotope Gini volume and normalized ME methods. The Gini volume is
computed based on the zonotope representation of the input-output data,
quantifying inequality in production contributions. The normalized ME
method estimates heterogeneity by calculating \(H_{max,t}\), the maximum
entropy of \(\log(Y)\) within clusters formed by K-means clustering of
\(\log(K)\) and \(\log(L)\) with 10 clusters. The entropy for each cluster
is derived as
\[
\log\left( y_{\max} - y_{\min} \right) + \log(\text{mean}(y)),
\]
weighted by the cluster's proportion of firms, to capture the dispersion
and central tendency of outputs. We perform 100 simulation runs to assess
the bias, variance, and accuracy of each method under low- and
high-heterogeneity conditions.

\subsection*{5.3 Results}
  \hspace*{2em}We compare the performance of the zonotope Gini volume and normalized
  maximum entropy (ME) methods in estimating industrial heterogeneity using
  the simulated high- and low-heterogeneity datasets. The results,
  summarized below, confirm that the Gini volume exhibits significant bias,
  particularly in high-heterogeneity settings, while the ME method provides
  more robust and consistent estimates. It is evident that the ME method
  correctly reflects the heterogeneity of datasets, while the Gini volume
  shows a slightly opposite trend.
  \begin{figure}[ht]
    \centering
    \includegraphics[width=0.95\textwidth]{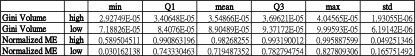}
    \caption{Comparison of Zonotope Gini Volume and Normalized Maximum Entropy (ME)  estimates across simulated datasets.}
  \end{figure}

\subsection*{6. Conclusion}
\hspace*{2em}In conclusion, this study addresses the limitations of traditional
methods in capturing the nuances of industrial heterogeneity by
introducing a maximum entropy approach for estimating production
functions and demonstrating how normalized entropy values can serve as
reliable metrics. By contrasting the entropy-based measures with a
zonotope-based Gini index using World Bank Enterprise Surveys data from
multiple countries, the research highlights the superiority of the
maximum entropy method in capturing functional heterogeneity often
missed by traditional techniques. The divergent trends observed between
the zonotope-based Gini value and normalized maximum entropy,
particularly in the case of Mexico's ISIC 10 industry,
underscore the importance of considering both quantitative and
qualitative aspects of heterogeneity.

\hspace*{2em}The illustrative example, though simplified, reveals that while firms
may converge in input-output quantities, technical change can lead to
greater diversity in production efficiency, a phenomenon more accurately
captured by the maximum entropy method. The study further strengthens
its claims through Monte Carlo simulations, which confirm the bias
inherent in the zonotope Gini volume method, particularly in
high-heterogeneity settings, and validate the robustness of the maximum
entropy method. Overall, this research offers a unified framework for
quantifying structural and functional differences across industries
using firm-level data, advancing both methodological and empirical
understanding of heterogeneity, which, in turn, is crucial for fostering
diverse economic ecosystems and understanding systemic risks within and
across sectors.

\section*{Appendix}
\subsection*{Summary of Y}
   \begin{center}
       \includegraphics[width=0.9\textwidth]{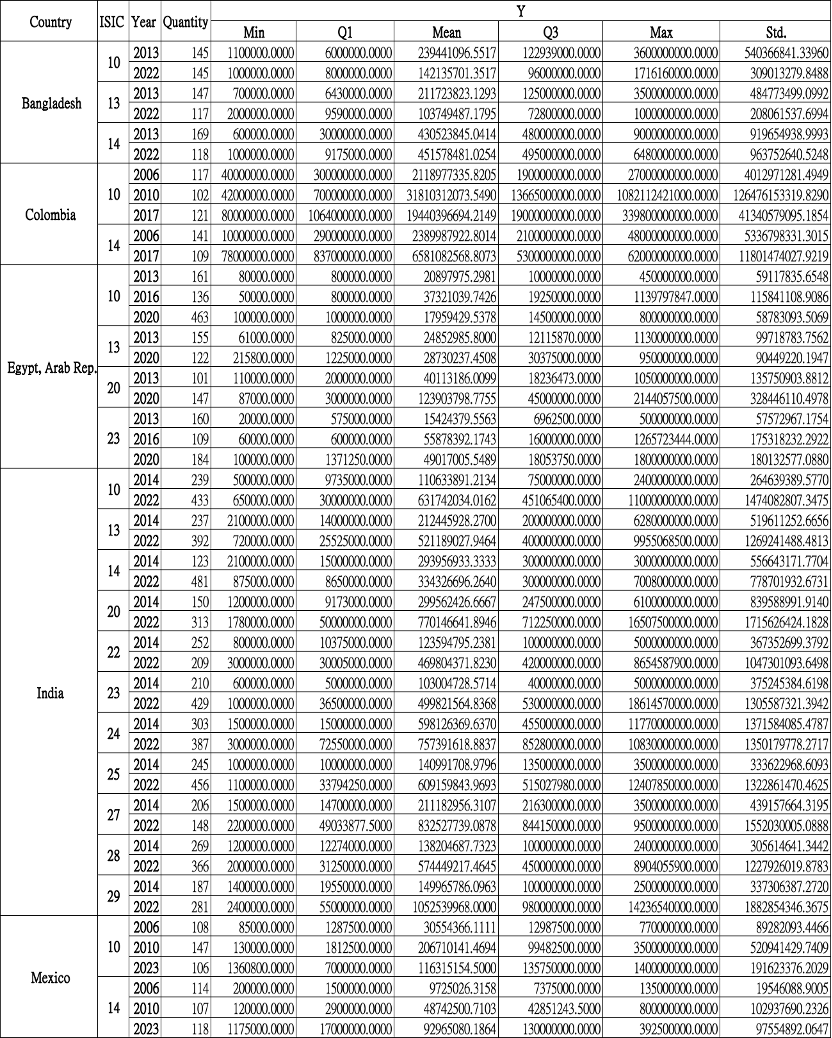}
   \end{center}
\subsection*{Summary of K}
    \begin{center}
        \includegraphics[width=0.9\textwidth]{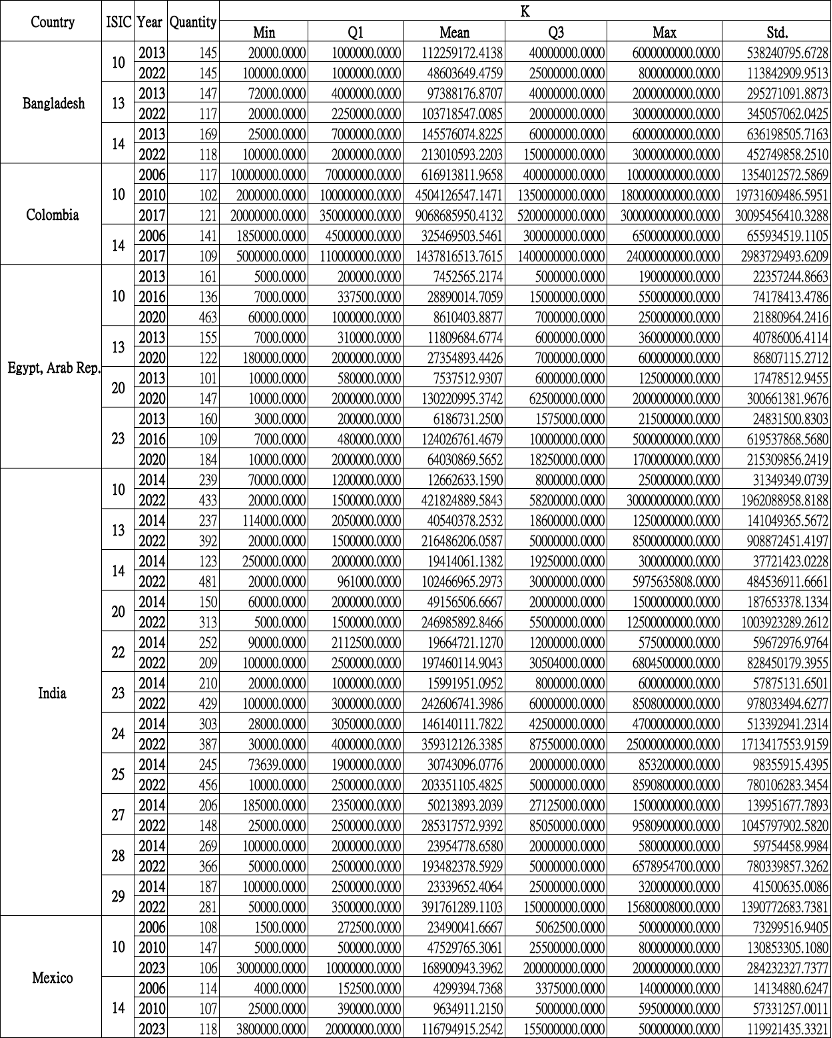}
    \end{center}
\subsection*{Summary of L}
    \begin{center}
        \includegraphics[width=0.9\textwidth]{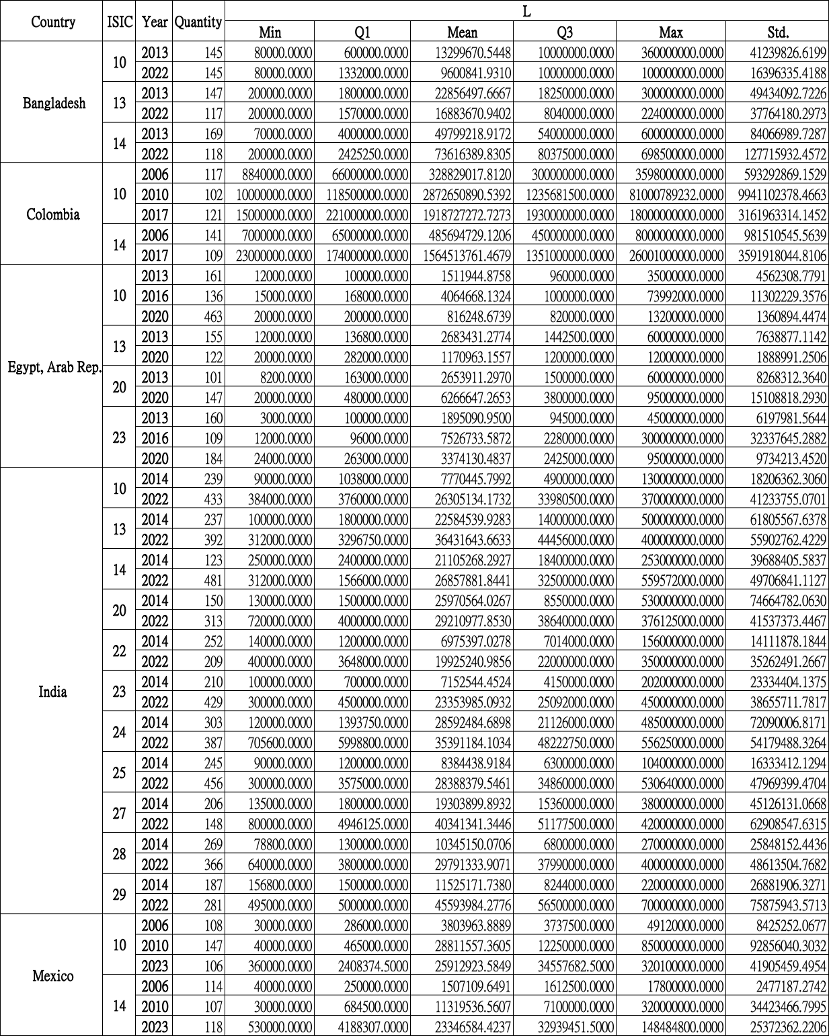}
    \end{center}
\end{document}